\pdfoutput=1
\documentclass[11pt,a4paper,twoside,openright,closeany,textany]{book}
\usepackage{makeidx}
\usepackage{graphicx}
\usepackage{paralist}
\usepackage{caption} % to set figure caption size
\usepackage{amsmath}
\usepackage{amssymb}
\usepackage{mathtools}
\usepackage{fancyhdr}
\usepackage{multicol}
\usepackage{multirow}

\usepackage{mdframed}% http://ctan.org/pkg/mdframed
\usepackage[usenames,dvipsnames,svgnames,table]{xcolor}% http://ctan.org/pkg/xcolor
\usepackage{array}
\usepackage[%
  colorlinks=true,%
  linkcolor=black,%
  urlcolor=black,%
  citecolor=black%
]{hyperref}
\usepackage{nameref} % to enable reference to sec/chap by name
\usepackage{ifthen} % include logic
\usepackage[font={small,sl}]{caption}
\usepackage[authoryear]{natbib}
\renewcommand\bibname{References}
\newcommand{\mychapbib}{% use this at end of each chapter
  \addcontentsline{toc}{section}{\bibname}
  \bibliographystyle{natbib}
  \bibliography{strucbioinf}
}
\usepackage[colorinlistoftodos]{todonotes}
\usepackage{letltxmacro}

\usepackage{tocstyle} % to set toc raggedright
\usetocstyle{standard}
\settocfeature{raggedhook}{\raggedright}
\newcommand{\bibfont}{\small\raggedright}
\def\cite{\citep}

\LetLtxMacro{\oldTodo}{\todo}
\renewcommand{\todo}[2][]{\oldTodo[#1]{TODO: #2}}

\usepackage[normalem]{ulem}
\newcommand\inwish[1]{\oldTodo[inline,color=SkyBlue]{WISH: #1}}

\newboolean{onechapter}

\newcommand{\AF}[1][~]{K.\@#1Anton#1Feenstra}
\newcommand{\SA}[1][~]{Sanne#1Abeln}

\newcommand{\AJ}[1][~]{Annika#1Jacobsen}
\newcommand{\HM}[1][~]{Halima#1Mouhib}

\newcommand{\BS}[1][~]{Bas#1Stringer}

\newcommand{\EvD}[1][~]{Erik#1van#1Dijk}
\newcommand{\NB}[1][~]{\mbox{Nicola}#1\mbox{Bonzanni}}
\newcommand{\OI}[1][~]{Olga#1Ivanova}

\newcommand{\RH}[1][~]{\mbox{Reza}#1\mbox{Haydarlou}}

\newcommand{\JG}[1][~]{\mbox{Jose}#1\mbox{Gavald\'a-Garc\'ia}}

\newcommand{\LH}[1][~]{\mbox{Laura}#1\mbox{Hoekstra}}

\newcommand{\orcid}[1]{\href{https://orcid.org/#1}{\raisebox{-0.7ex}{\protect\includegraphics[height=3ex]{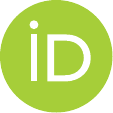}}}}
\definecolor{idgreen}{RGB}{166 206 57}
\newcommand{\mailid}[1]{\href{mailto:#1}{\raisebox{-0.3ex}{\color{idgreen}\textsf{\textbf{\Large \protect@}}}}}

\newcommand{\AFid}{\orcid{0000-0001-6755-9667}}
\newcommand{\SAid}{\orcid{0000-0002-2779-7174}}
\newcommand{\HMid}{\orcid{0000-0001-5031-3468}}
\newcommand{\JGid}{\orcid{0000-0001-6431-3442}}

\newcommand{\OIid}{\orcid{0000-0002-9111-4593}}
\newcommand{\RHid}{\orcid{0000-0003-4138-7179}}

\newcommand{\BSid}{\orcid{0000-0001-7792-9385}}
\newcommand{\AJid}{\orcid{0000-0003-4818-2360}}
\newcommand{\EvDid}{\orcid{0000-0002-6272-2039}}

\newcommand{\LHid}{\orcid{0009-0003-4339-2470}}

\newcommand{\NBid}{\mailid{nickola80@gmail.com}}

\newcommand{\ACtxt}{Wrote the text}
\newcommand{\ACfig}{Created figures}
\newcommand{\ACref}{Review of current literature}
\newcommand{\ACeds}{Editorial responsibility}
\newcommand{\ACproof}{Critical proofreading}

\newcommand{\Angs}[1][~]{\text{\normalfont\AA}}

\renewcommand{\and}{\quad}

\newcommand{\figlab}[1]{\textsf{\textbf{\Large #1}}}
\newcommand{\pdbref}[1]{\href{http://www.rcsb.org/pdb/explore.do?structureId=#1}{PDB:#1}}
\newcommand{\arxiv}[2][UNDEFINED]{\href{https://arxiv.org/abs/#2}{\ifthenelse{\equal{#1}{UNDEFINED}}{arxiv.org/abs/#2}{#1}}}

\newcommand{\figref}[2][]{\hyperref[fig:#2]{Figure\@~\ref*{fig:#2}#1}}
\newcommand{\tabref}[1]{\hyperref[tab:#1]{Table \ref*{tab:#1}}}
\renewcommand{\eqref}[2][]{\hyperref[eq:#2]{Equation#1\@~\ref*{eq:#2}}}
\newcommand{\panelref}[2][]{%
    \ifthenelse{\boolean{onechapter}}{%
        \hyperref[panel:#2]{Panel\@~``\nameref{panel:#2}#1''}%
    }{%
        \hyperref[panel:#2]{Panel\@~\ref*{panel:#2}#1}%
    }%
}
\newcommand{\secref}[2][n]{%
    \hyperref[sec:#2]{%
        \ifthenelse{\equal{#1}{n} }{Section\@~\ref*{sec:#2}}{}% just number
        \ifthenelse{\equal{#1}{nn}}{Section\@~\ref*{sec:#2} ``\nameref{sec:#2}''}{}% nm & nr
        \ifthenelse{\equal{#1}{N} }{``\nameref{sec:#2}''}{}% just quoted name
        \ifthenelse{\equal{#1}{NN} }{\nameref{sec:#2}}{}% just name
    }%
}
\newcommand{\chref}[2][n]{%
    \ifthenelse{\boolean{onechapter}}{%
        \ifthenelse{\equal{#2}{ChPref}     }{\arxiv[Chapter ``\nameref*{ch:#2}'']{1801.09442}}{}%
        \ifthenelse{\equal{#2}{ChIntroPS}  }{\arxiv[Chapter ``\nameref*{ch:#2}'']{1801.09442}}{}%
        \ifthenelse{\equal{#2}{ChDetVal}   }{\arxiv[Chapter ``\nameref*{ch:#2}'']{2108.02706}}{}%
        \ifthenelse{\equal{#2}{ChStrucAli} }{\arxiv[Chapter ``\nameref*{ch:#2}'']{1801.09442}}{}%
        \ifthenelse{\equal{#2}{ChDBClass}  }{\arxiv[Chapter ``\nameref*{ch:#2}'']{1801.09442}}{}%
        \ifthenelse{\equal{#2}{ChFunc}     }{\arxiv[Chapter ``\nameref*{ch:#2}'']{1801.09442}}{}%
        \ifthenelse{\equal{#2}{ChIntroPred}}{\arxiv[Chapter ``\nameref*{ch:#2}'']{1712.00407}}{}%
        \ifthenelse{\equal{#2}{ChHomMod}   }{\arxiv[Chapter ``\nameref*{ch:#2}'']{1712.00425}}{}%
        \ifthenelse{\equal{#2}{ChSSPred}   }{\arxiv[Chapter ``\nameref*{ch:#2}'']{1801.09442}}{}%
        \ifthenelse{\equal{#2}{ChFuncPred} }{\arxiv[Chapter ``\nameref*{ch:#2}'']{1801.09442}}{}%
        \ifthenelse{\equal{#2}{ChIntroDyn} }{\arxiv[Chapter ``\nameref*{ch:#2}'']{1801.09442}}{}%
        \ifthenelse{\equal{#2}{ChThermo}   }{\arxiv[Chapter ``\nameref*{ch:#2}'']{1801.09442}}{}%
        \ifthenelse{\equal{#2}{ChMD}       }{\arxiv[Chapter ``\nameref*{ch:#2}'']{1801.09442}}{}%
        \ifthenelse{\equal{#2}{ChMC}       }{\arxiv[Chapter ``\nameref*{ch:#2}'']{1801.09442}}{}%
    }{% else
    \hyperref[ch:#2]{%
        \ifthenelse{\equal{#1}{n} }{Chapter \ref*{ch:#2}}{}% just number
        \ifthenelse{\equal{#1}{nn}}{Chapter \ref*{ch:#2} ``\nameref{ch:#2}''}{}% name & number
        \ifthenelse{\equal{#1}{N} }{``\nameref{ch:#2}''}{}% just name
      }%
  }%
}
\newcommand{\chrefname}[1]{\hyperref[ch:#1]{Chapter \ref*{ch:#1} ``\nameref{ch:#1}''}}
\newcommand{\partref}[1]{\hyperref[#1]{Part \ref*{#1}}}
\newcommand{\appref}[1]{\hyperref[app:#1]{Appendix \ref*{app:#1}}}

\newcommand{\figsource}[1]{\protect\footnote{Figure source location: \url{#1}}}

\newenvironment{penum}[1][\itshape i)\upshape]
{\begin{inparaenum}[#1]} {\end{inparaenum}}
\newenvironment{cenum}[1][\itshape i)\upshape\ ]
{\begin{compactenum}[#1]} {\end{compactenum}}

\renewcommand{\arraystretch}{1.3}

\makeatletter \@beginparpenalty=5000 \makeatother

\newenvironment{panel}[1][]{
  \begin{figure}[htb]
    \begin{mdframed}[%
        outerlinewidth=0,%
        linecolor=CornflowerBlue!30,%
        backgroundcolor=CornflowerBlue!30,%
        innerleftmargin=14,%
        innerrightmargin=14,%
      ]
      \ifthenelse{\equal{#1}{}}{}{% only if optional arg not empty
        \stepcounter{panel}
		\subsection*{#1} % without panel numbers
      }
}{%
    \end{mdframed}
  \end{figure}
}

\newenvironment{bgreading}[1][]{
  \begin{mdframed}[%
      outerlinewidth=0,%
      linecolor=CornflowerBlue!30,%
      backgroundcolor=CornflowerBlue!30,%
      innerleftmargin=14,%
      innerrightmargin=14,%
    ]
	\ifthenelse{\equal{#1}{}}{}{% only if optional arg not empty
        \stepcounter{panel}
    	\subsection*{#1} % without panel numbers
    }
}{%
  \end{mdframed}
}

\usepackage{listings}
\definecolor{backcolour}{rgb}{0.95,0.95,0.92}
\definecolor{codegreen}{rgb}{0,0.6,0}
\definecolor{codegray}{rgb}{0.5,0.5,0.5}
\definecolor{codered}{rgb}{0.8,0,0.0}
\definecolor{codeblue}{rgb}{0.0,0,0.8}
\lstdefinestyle{codeStyle}{
    backgroundcolor=\color{backcolour},   
    commentstyle=\color{codegreen},
    keywordstyle=\color{codeblue},
    numberstyle=\tiny\color{codegray},
    stringstyle=\color{codegray},
    numbers=left,                    
    tabsize=2
} 
\makeindex

\begin{document}

\setboolean{onechapter}{true}

% page head and foots (feet):
\pagestyle{fancy}
\lhead[\small\thepage]{\small\sf\nouppercase\rightmark}
\rhead[\small\sf\nouppercase\leftmark]{\small\thepage}
\newcommand{\innerfoot}{\footnotesize{\sf{\copyright} Feenstra \& Abeln}, 2014-2023}
\newcommand{\outerfoot}{\footnotesize \sf Intro Prot Struc Bioinf}
\lfoot[\outerfoot]{\innerfoot}
\cfoot{}
\rfoot[\innerfoot]{\outerfoot}
\renewcommand{\footrulewidth}{\headrulewidth}

\mainmatter
\setcounter{chapter}{0}
\chapterauthor{\AJ~\AJid \and \EvD~\EvDid \and \HM~\HMid \and \BS~\BSid \and \OI~\OIid \and \JG~\JGid \and \LH*~\LHid \and \AF*~\AFid \and \SA*~\SAid}
\chapterfootnote{* editorial responsability}
\chapterfigure{\includegraphics[width=0.5\linewidth]{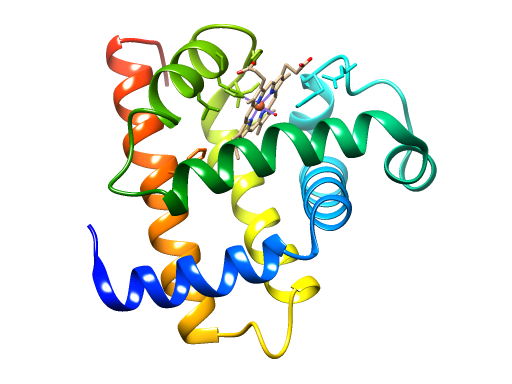}}
\chapter{Introduction to Protein Structure}
\label{ch:ChIntroPS}

\ifthenelse{\boolean{onechapter}}{\tableofcontents\newpage}{} 

Within the living cell, protein molecules perform specific functions, typically by interacting with other proteins, DNA, RNA or small molecules. They take on a specific three dimensional structure, or in some cases, an ensemble of three dimensional structures. It is this three dimensional structure that allows the protein to function within the cell. This structure is with high specificity encoded by its amino acid sequence; the precise amino acid sequence of a protein is in turn encoded by the genes of an organism. Hence, the understanding of a protein's function is tightly coupled to its three dimensional structure. 

The current state of scientific understanding allows us to comprehend how the gene sequence encoded by the DNA is transcribed into RNA, and in its turn translated into amino acid sequence. However, experiments to determine protein structures and protein structural ensembles are difficult and laborious; we will come back to that in detail in \chref{ChDetVal}. Recently, deep learning models like AlphaFold2, trained on existing protein structure data, have achieved success in predicting protein structures from sequences. However, simulating the transition from a protein sequence to its folded structure computationally is still challenging for moderately sized proteins. As a result, structural bioinformatics faces unresolved problems or, alternatively, presents exciting scientific challenges.

Before going into protein structure analysis and prediction, and protein folding and dynamics, we will first give a brief introduction into the basics of protein structures. This is deliberately kept short and shallow. The excellent book ``Introduction to Protein Structure'' by \citet{BrandenTooze} provides a much more in-depth introduction into this exciting field.

\begin{figure}
\includegraphics[width = \linewidth]{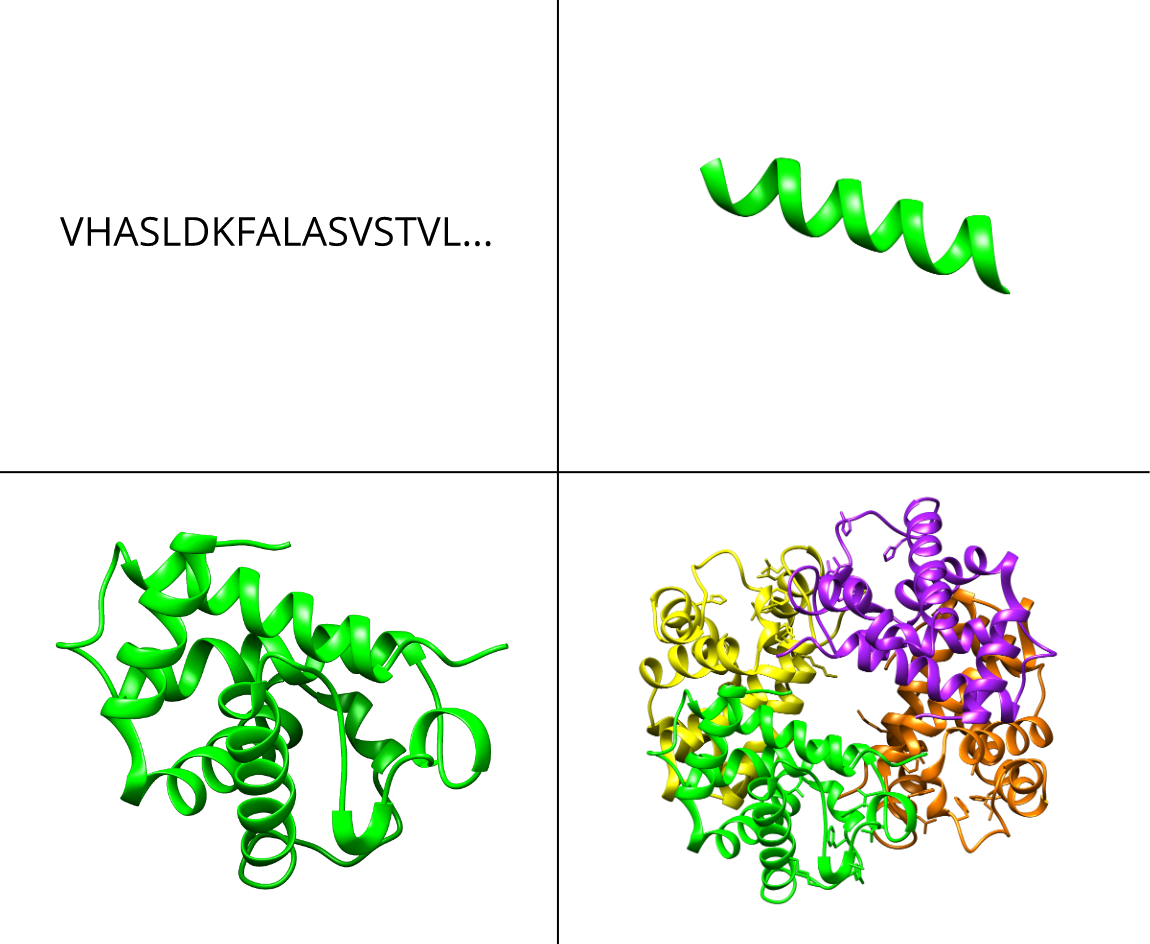}
\caption{\textbf{Levels of protein structure.} Top-left: Primary structure, given as polypeptide sequence in the one-letter code of amino acids. Top-right: Secondary structure, example of an alpha helix. Bottom-left: Tertiary structure, structure of one of the monomers of hemoglobin. Bottom-right: Full structure of Human hemoglobin, 4 chains make the whole structure (\pdbref{1BIJ}). Ribbon representation obtained with UCSF-chimera \cite{Pettersen2004}.}
\label{fig:ChIntroPS:structures_pstq}
\end{figure}

\section{Protein structure basics}
A protein structure may be described at four levels as depicted in \figref{ChIntroPS:structures_pstq}:
\begin{compactenum}
\item[\textbf{The primary structure}] is simply the sequence of amino acids that make up the protein polypeptide chain.

\item[\textbf{Secondary structure}] describes the organisation of this chain into regular $\alpha$-helices and $\beta$-strands and anything else, called `coil' or loop.

\item[\textbf{Tertiary structure}] is the three dimensional arrangement -- or topology -- of the protein chain; it defines its overall shape.

\item[\textbf{Quaternary structure}] is the (three dimensional) organisation of the protein chain in context of the proteins and molecules it interacts with; i.e.\@ the configurational ensemble multiple molecules adopt when binding to each other, forming macro-molecular complexes.
\end{compactenum}

\subsection{Primary structure}

There are 20 naturally occurring amino acids that constitute the building blocks of proteins. 
Amino acids are linked together by a \textbf {peptide-bond} between the carbonyl Carbon (C=O) of the preceding residue and the amide Nitrogen (NH) of the next residue in its primary sequence. This is why proteins are also referred to as ``polypeptides''. Note that for each amino acid type, this part of the chemical structure is identical; it is also referred to as the \emph{backbone}. The \emph{sidechains} branch out from the central Carbon atom (C$\alpha$) in the backbone. Unlike the backbone, sidechains are chemically different between the different amino acid types; see \panelref{ChIntroPS:aas} for more detail. We can view the primary protein structure as a chain with 20 different colours of beads that all have different properties.

The amino acid type in the polypeptide chain is encoded by codons (sequences of three nucleotides) in the DNA sequence of a gene. After transcription, the translation from RNA codons into amino acids occurs at the ribosome. The exact sequence of the codon determines the amino acid, or may indicate the start or end of a peptide chain. This codon table is universal across all species, although several microorganisms are known that use a (slightly) different table. The translation mechanism, including the codon table, the tRNA and the aminoacyl-transferases, is beyond the scope of this book. 

Roughly speaking, there are three important classes of amino acids: 
\begin{penum}
\item hydrophobic,
\item charged, and
\item polar%
\end{penum}.
These classes are based on their interaction properties with respect to water. Hydrophobic residues do not interact with water, whereas polar and charged residues do make contact with water favourably. Later in this chapter, we will see the importance of the difference between hydrophobic and polar amino acids for protein folding. The \panelref{ChIntroPS:aas} gives more background detail on the chemical characteristics of those amino acids.

\begin{bgreading}[Amino acids, residues, and the peptide bond]
\label{panel:ChIntroPS:aas}
There are 20 naturally occurring amino acids that constitute the building blocks of proteins, shown in \panelref{naturalaa} below. (Chemically speaking many, many more types of amino acids are possible). To build up a protein, amino acids react under the loss of water to form an extremely stable peptide bond. The figure shows the formation of the peptide bond between two alanine amino acids:  

\centerline{\includegraphics[width=0.6\linewidth]{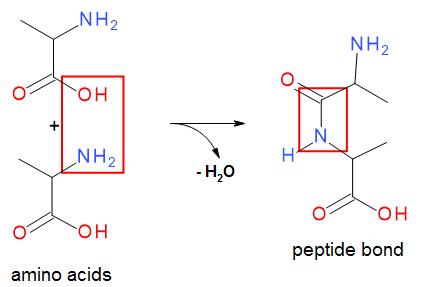}} 

\noindent
Within proteins, amino acids differ in the sidechain part, the backbone of the protein, i.e., (NH--C$\alpha$--C=O) is repetitive. Note that each amino acid can be referred to using a three or one letter code: here Ala or A for alanine. In \panelref{naturalaa}, amino acid residues are categorized into \textbf{charged} (positive/negative), \textbf{polar} (not charged), and \textbf{hydrophobic} (or aliphatic) amino acids. 

\begin{cenum}
\item[\bf Hydrophobic/aliphatic/apolar] amino acids consist of only Carbon (and for Cysteine, Sulphur) atoms in the sidechain. 
\item[\bf Aromatic residues] all have a regular six- or five-sided ring (tryptophan has both) consisting of mostly carbon atoms. Tyrosine is aromatic but due to a hydroxyl group also polar. Tryptophan does contain a nitrogen atom but is considered hydrophobic. 
\item[\bf Polar and charged] residues all have nitrogen and/or oxygen atoms in the sidechain. Charged residues are also considered polar, and both are hydrophilic.
\item[\bf Small, medium, large:] A further subdivision of the hydrophobic amino acids can be made into small (glycine, alanine), medium (valine, leucine, isoleucine), large (methionine) and aromatic/ring (proline, which is also medium size, and phenylalanine and tryptophan which are also large). 
Of the charged amino acids, lysine and arginine are considered large, and among the polar ones, serine and threonine are small. 
\end{cenum}
Finally, the backbone contains N and O and is therefore always polar. In the polar and charged residues, oxygen is always negative and nitrogen always positive.

Before moving on, it should be noted that biologists and bioinformaticians often use the terms `amino acid' and `residue' equivalently. However, `residue' is more general and can also refer to e.g.\@ a nucleotide in DNA or RNA. To be a bit more precise, for chemists, the amino acid is the free molecule, and it is called `amino acid residue' only when part of a protein. 

\subsubsection{The 20 natural amino acid residues}
\label{panel:naturalaa}
\centerline{\includegraphics[width=0.95\linewidth]{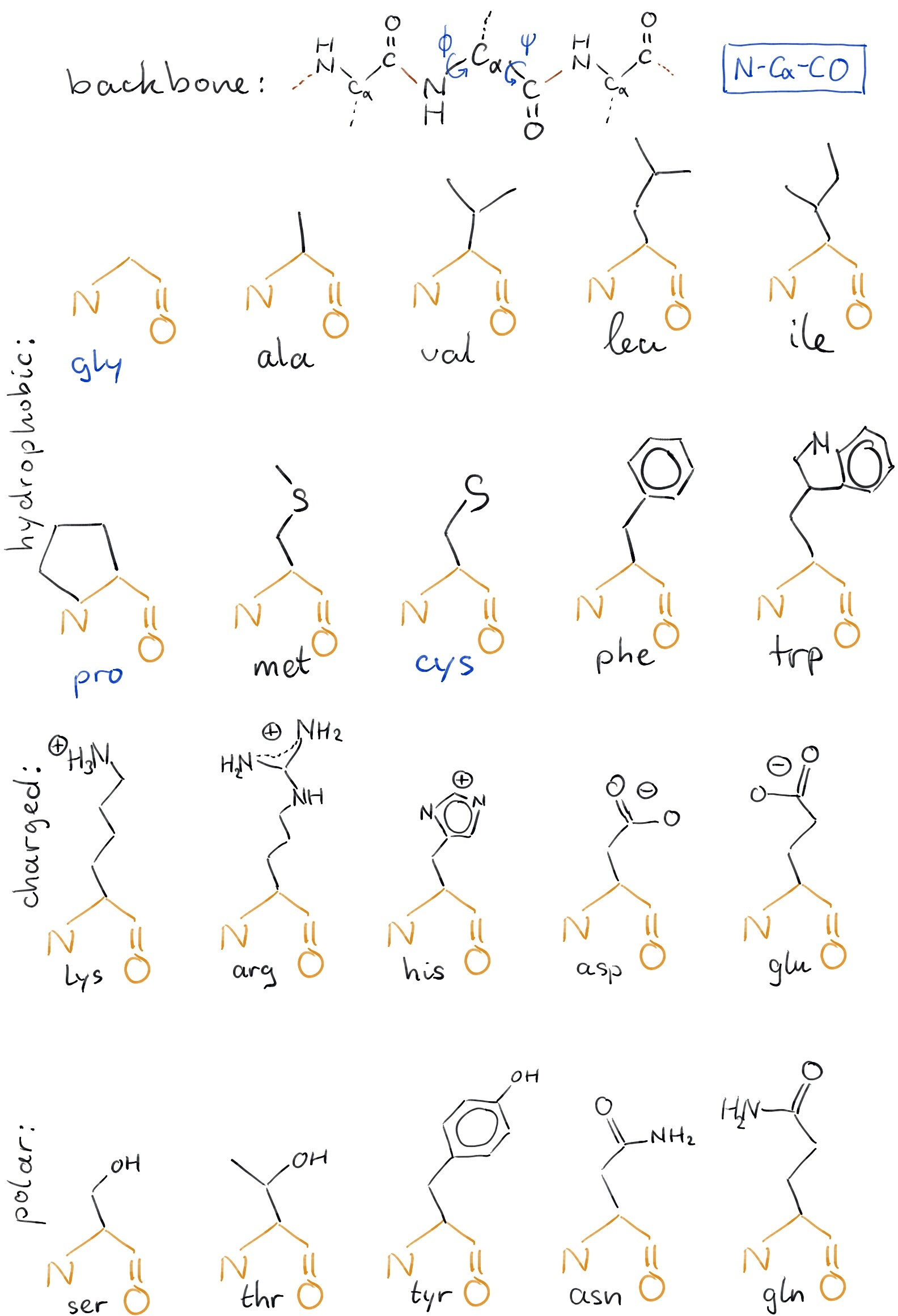}}

In addition to these broad categories of amino acids, there are a few special ones (labelled in blue in the figure): 
\begin{compactenum}
\item[\bf Cysteine] contains a sulphur atom. When two cysteines are close in the structure, the two sulphur atoms will form a covalent bond of similar strength to the other bonds within the amino acids. These are much stronger than hydrogen bonds (see next 1.1.2 for hydrogen bonds). 

\item[\bf Proline] contains a ring that loops from the C$\alpha$ back to the backbone nitrogen. This makes the backbone of the proline much less flexible than for other residues; proline often terminates helices or otherwise induces a kink, and proline is used to make a loop containing sharp turns. More about this below in the \panelref{ChIntroPS:omega}.

\item[\bf Glycine] has the smallest possible sidechain: only a single hydrogen atom (which is much smaller than a carbon). Due to this, there is less steric hindrance around the C$\alpha$ and more flexibility in its Phi/Psi angles (more detail about phi/psi angles in \secref{ChIntroPS:phi-psi} below).
\end{compactenum}
\end{bgreading}

\begin{figure}[b]
\centerline{
  \includegraphics[width=0.8\linewidth]{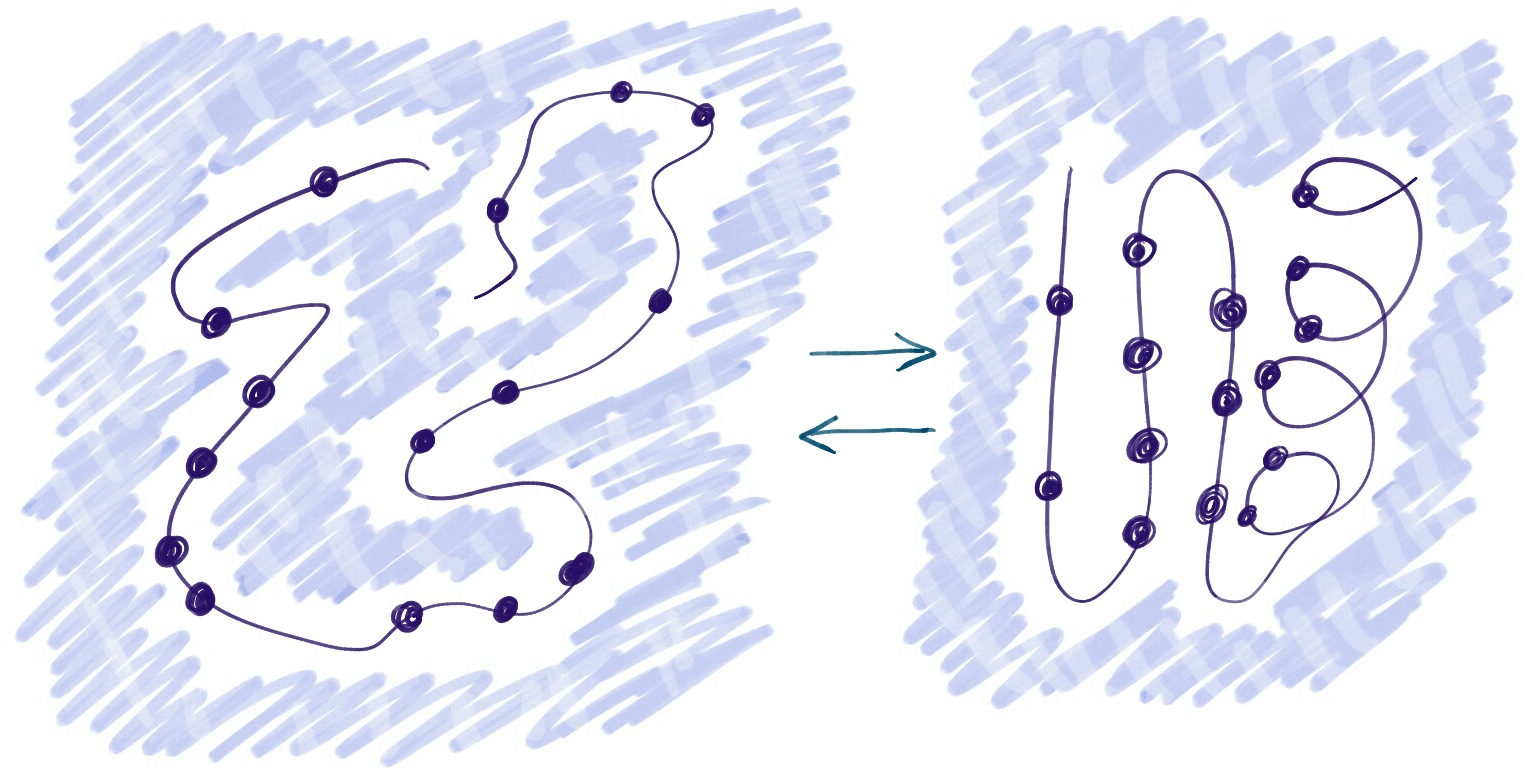}
}
\caption{Hydrophobic collapse as the first step in a protein folding from its unfolded state (on the left) to a folded state (right). Hydrophobic residues, shown as black spheres, will tend to minimize contact with water and therefore end up in the interior of the protein. Hydrophilic (polar and charged) residues are not drawn explicitly here, they form the rest of the backbone, between the black spheres.}
\label{fig:ChIntro-collapse}
\end{figure}

\begin{figure}
\centerline{\includegraphics[width=\linewidth]{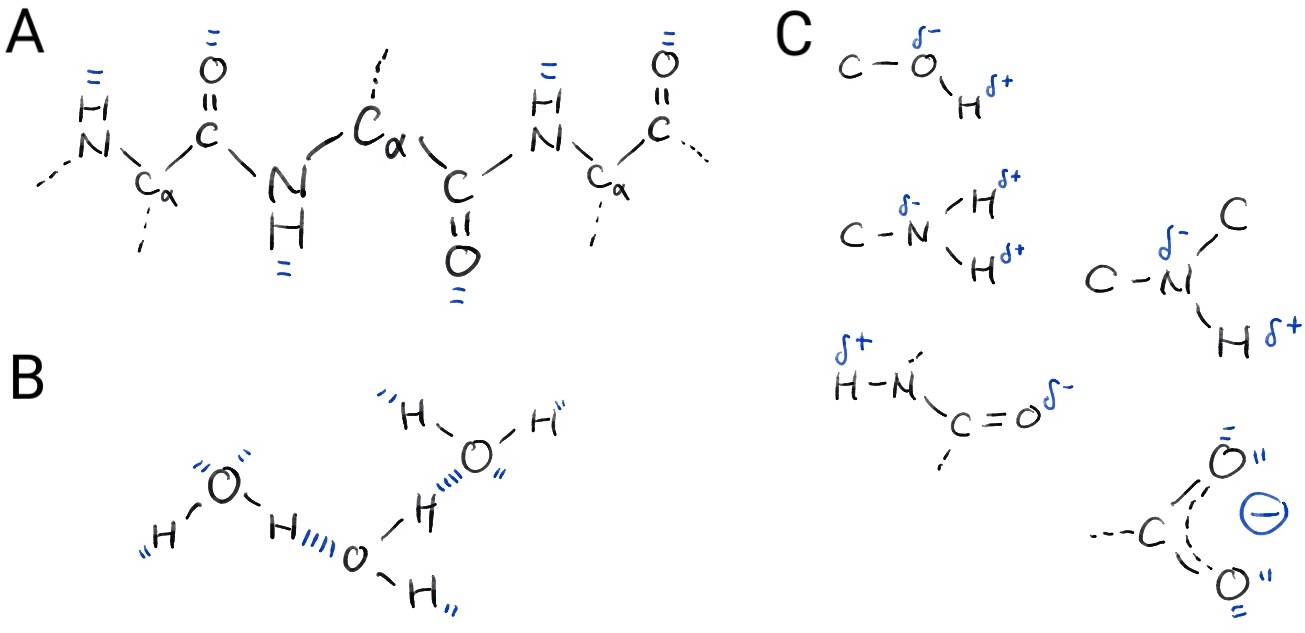}}
\caption{Hydrogen bonding in the backbone of the protein (A) and in water (B); hydrogen-bond forming groups are indicated with blue dashed lines. Hydrogen bonds are caused by atoms with slight negative charges ($\delta-$) being attracted to atoms with slightly positive charges ($\delta+$). In a protein, hydrogens on a nitrogen or oxygen are positive, oxygens and nitrogens themselves are negative (C).}
\label{fig:ChIntroPS-hbonds}
\end{figure}

\section{Secondary structure}

\begin{figure}
\centerline{\includegraphics[width=0.95\linewidth]{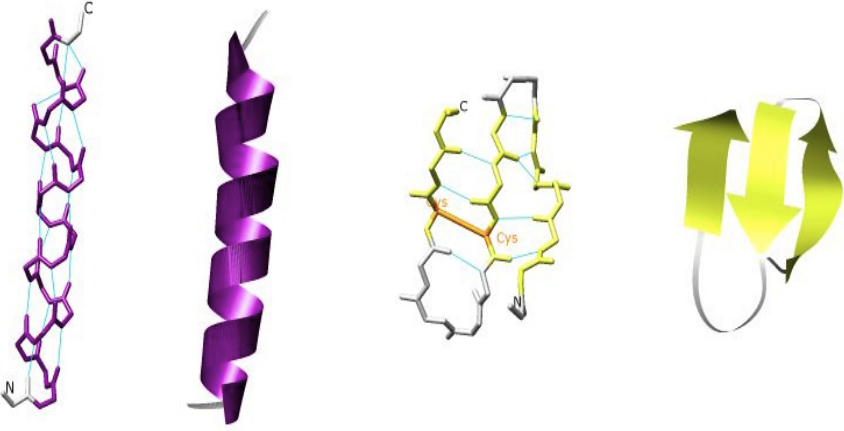}}
\caption{(Examples of $\alpha$-helical (left two) and $\beta$-strand (right two) structures made out of alanines (note the single atom in the sidechain) and a few cysteines (in the $\beta$-sheet). Both are shown in a `sticks' (left) and a typical `cartoon' (right) representation.}
\label{fig:ChIntroPS-helix-strand} %\remark{(SA: can be used)
\end{figure}

All amino acids have a common part, the backbone, as discussed in the previous section. The backbone can make regular structures due to their chemical properties, see also \figref{ChIntroPS-hbonds}. These regular backbone structures are called secondary structure. Examples of $\alpha$-helices and $\beta$-strands are shown in \figref{ChIntroPS-helix-strand}. Below we will introduce how backbone hydrogen bonding leads to secondary structure.

\subsection{Backbone hydrogen bonding}

Hydrogen bonds are a key part of the secondary and tertiary structure of a protein. Hydrogen bonding takes places between hydrogen atoms, with a slight positive charge and nitrogen or oxygen atoms with a slight negative charge. 

The attraction of opposite charges leads to strong H$\cdots$O interactions called \textbf{hydrogen-bonds}. The N--H and C=O parts of the backbone, can also make these polar hydrogen-bond interactions (see \figref{ChIntroPS-hbonds}), and so can the polar and charged sidechains (O--H, N--H, S--H and C=O groups as shown in \panelref{naturalaa}). Energetically, hydrogen-bonds are very favourable, and most hydrogen bond donors and acceptors in the sequence of the protein will therefore also make a hydrogen bond, in any stable structure.

There are two main ways in which the amino acid chain in proteins are structured so that all backbone hydrogen bonds in the hydrophobic interior can become satisfied: \emph{helix} or \emph{strand}. In the helical structure, repeated local hydrogen bonding occur, as shown in \figref{ChIntroPS-helix-strand}. This is called the $\alpha$-helix secondary structure type. An $\alpha$-helix only leaves unsatisfied hydrogen bonding capacity at the ends of a helix, these ends are thus usually found at the surface of a protein. In the strand structure, two stretches of sequence are adjacent in the structure, and hydrogen bonds occur `laterally' between adjacent strands. This is the `$\beta$-sheet' secondary structure. It leaves unsatisfied hydrogen bond capacity at the first and last strands of a sheet, called the `edge' strands, which like helix ends are mostly found at the protein surface. 

\subsection{$\alpha$-helices}

Helical secondary structures are characterized by repeated, local hydrogen bonding between the backbone amide group of one residue, and the carbonyl group of a subsequent residue, as shown in \figref{ChIntroPS-helix-strand}. The most common of these structures is the $\alpha$-helix, where the bond is formed between residues $i$ and $i+4$. Helices can be anywhere from 4 to 40 residues long, with an average length of $\sim$10 residues, or about 3 turns. 

This periodicity of $\alpha$-helices can typically also be observed in the sequence. Within a protein structure a helix typically has a solvent exposed and a buried side; this will lead to hydrophilic residues tending to point outside towards the solvent, and hydrophobic ones tending towards the inside of the protein. The side sticking into the core are typically tightly packed together, also referred to as helix packing.
Helical structures are generally considered easier to predict from sequence due to this periodicity in the sequence, we will come back to this in \chref{ChSSPred}.
Please refer to the \panelref{ChIntroPS:helices} below for further details on helices.

\begin{bgreading}[Helices]
\label{panel:ChIntroPS:helices}
The $\alpha$-helix is the most common helix secondary structure type, where the bond is formed between residues $i$ and $i+4$. So-called $3/10$- and $\pi$-helices, where the bond connects residues $i \rightarrow i+3$ and $i \rightarrow i+5$ respectively, are less common. 
In the helical structure, the carbonyl group (C=O) is oriented along the direction of the sequence, while the amide group (NH) points in the opposite direction (\figref[a]{ChIntroPS-helix-strand}). This arrangement results in a tightly packed helix with minimal internal space. 
\end{bgreading}

\begin{figure}
\centerline{\begin{tabular}{cc@{}c}
\multirow2*{\includegraphics[width=0.5\linewidth]{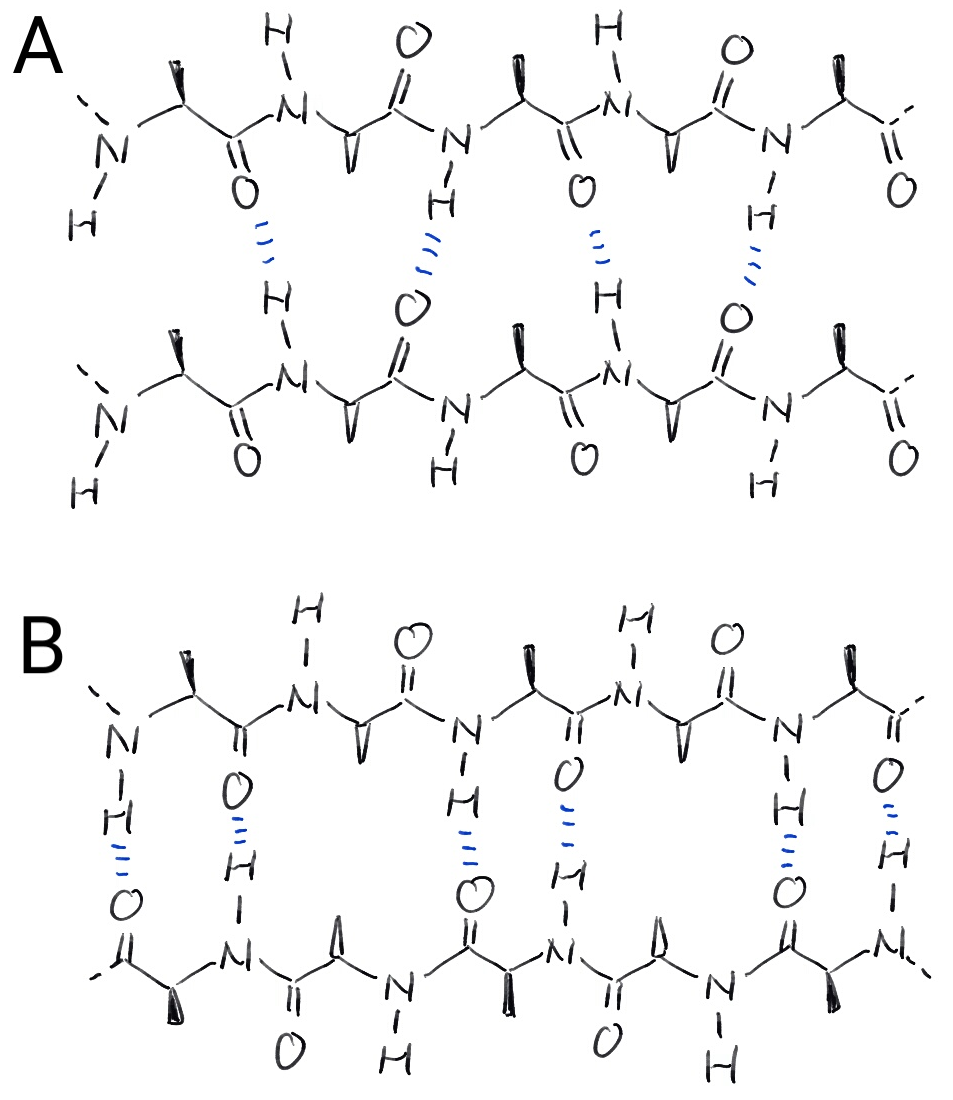}} &
\figlab{C} & 
\raisebox{-0.9\height}{
  \includegraphics[height=0.25\linewidth]{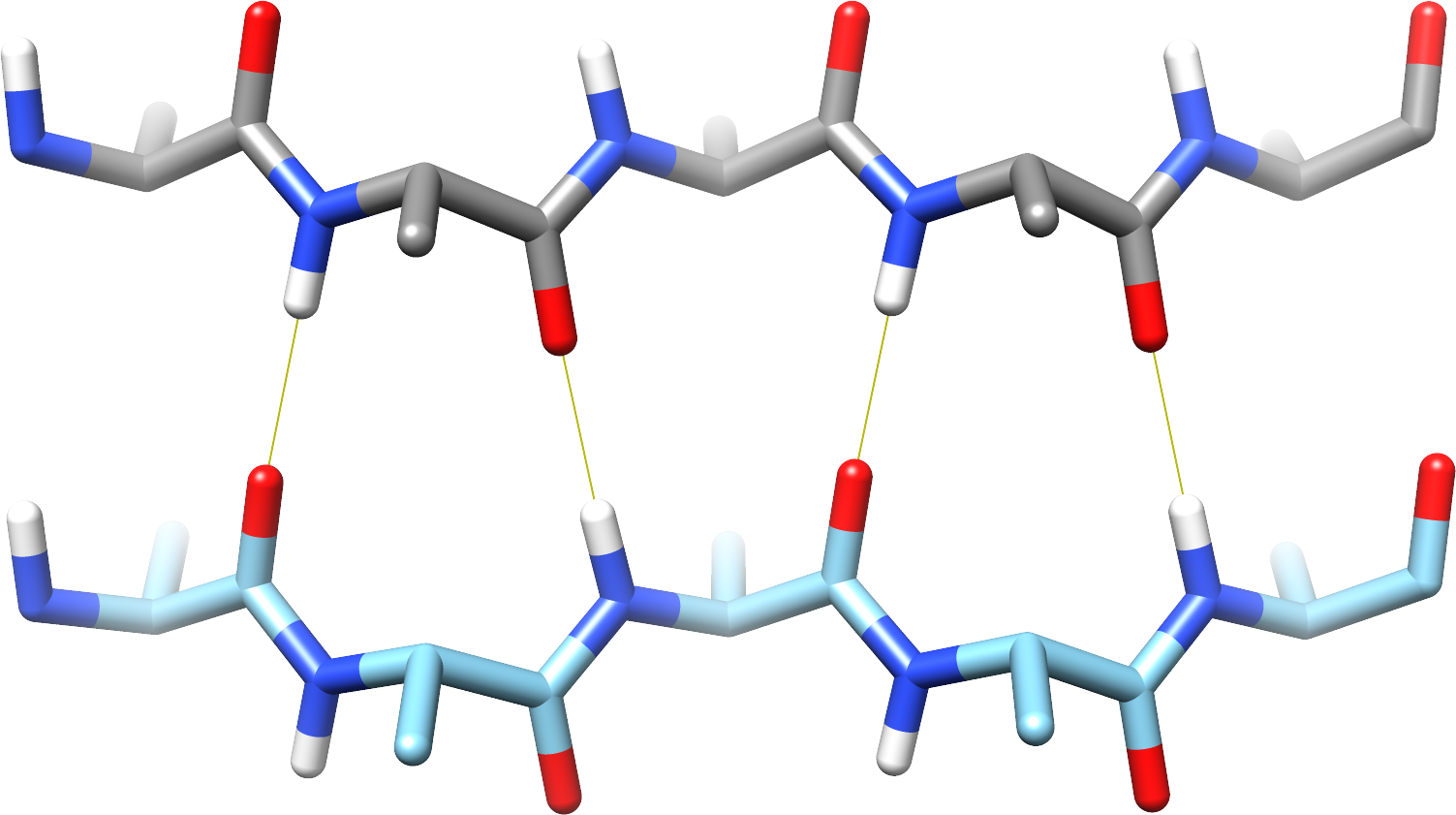}} \\
 & & \\
& \figlab{D} & 
\raisebox{-0.9\height}{
  \includegraphics[height=0.25\linewidth]{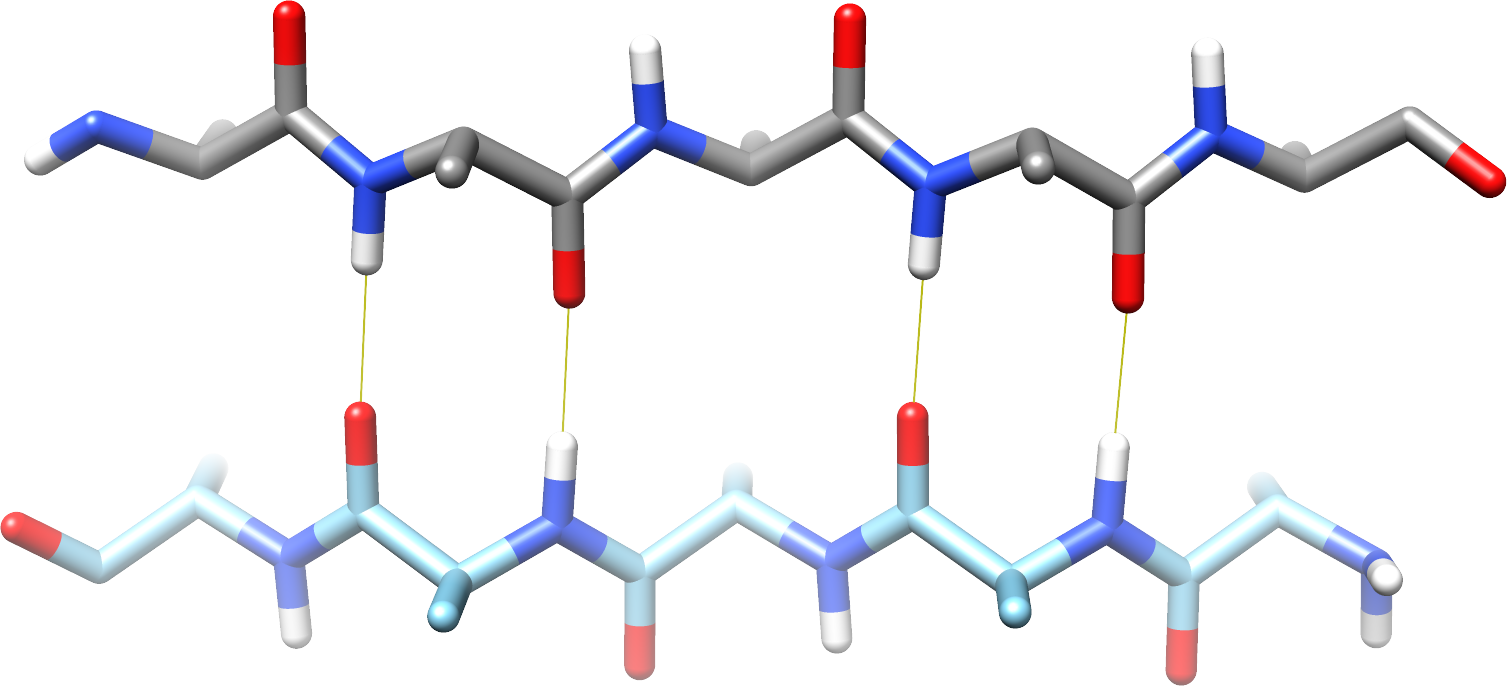}} \\
\end{tabular}}
\caption{Two types of $\beta$-sheet, schematically: a) parallel, and b) anti-parallel; and in three-dimensions: c) parallel and d) anti-parallel (ideal geometries, generated using Chimera).}
\label{fig:par_vs_antipar}
\end{figure}
\begin{figure}
\centerline{
  \figlab{A}\includegraphics[width=0.28\linewidth]{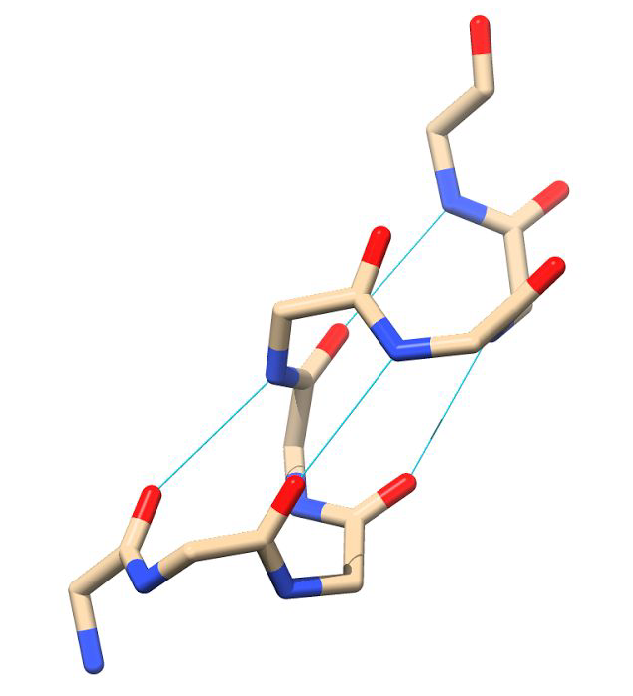}
  \figlab{B}\includegraphics[width=0.25\linewidth]{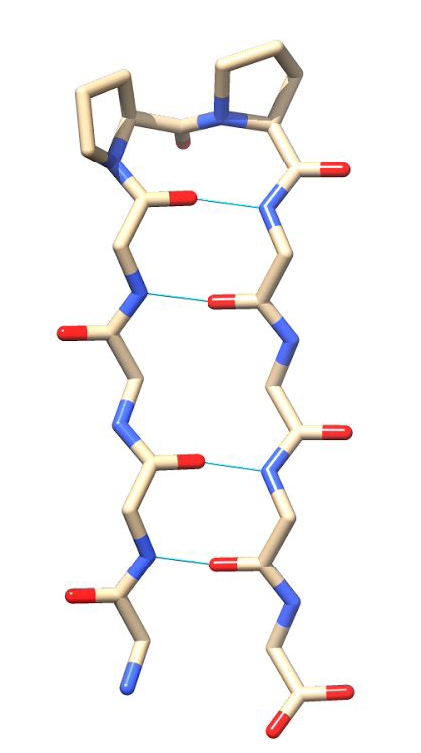}
  \figlab{C}\includegraphics[width=0.3\linewidth]{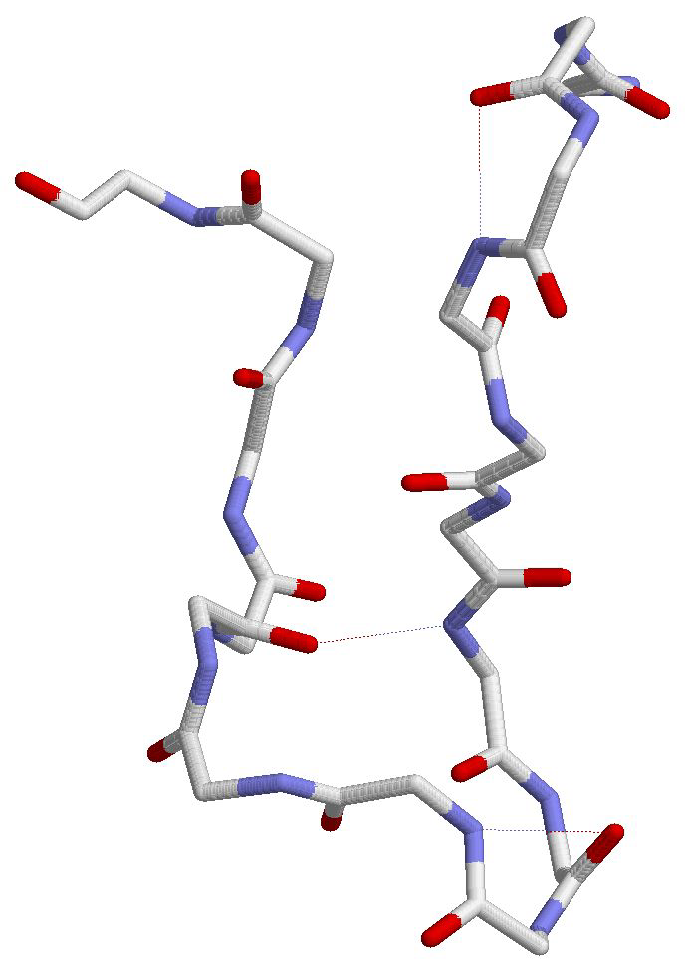}
}
\caption{Details of $\alpha$-helix (A), $\beta$-strand (B) and coil (C). Note how regular patterns of hydrogen bonds (thin lines) stabilize $\alpha$-helix (A) and $\beta$-strand (B), but not coil (C). Also note how the hydrogen bonds in the $\alpha$-helix (A) point along the helix axis; the hydrogen bonds go from the hydrogen atom (which is not shown) on the nitrogen (blue) to the oxygen (red), pointing 'backwards' along the direction of the protein chain, which runs from bottom left to top right (A).}
\label{fig:ChIntroPS-ss-detail} 
\end{figure}
\begin{panel}[Strands and sheets]
\label{panel:ChIntroPS:sheets}
 Due to the chirality of amino acids, $\beta$-sheets are often twisted or pleated in a right-hand turn. Simple configurations of $\beta$-sheets include the commonly found hairpin and so-called psi-loop motif, whereas larger sheets can assume complex formations like $\beta$-barrels or $\beta$-propellers.

When the connecting sequence between two strands is a small loop (a $\beta$-loop-$\beta$ motif, often referred to as $\beta$-hairpin) \figref[b]{ChIntroPS-ss-detail}, the sequence distance can be as low as 3. However, the sequence distance between two strands in a beta-sheet can be much larger. An extreme case occurs when a whole protein domain is in between the two strands, which might be hundreds of residues (more on protein domains later in \chref[nn]{ChDBClass}). Therefore, no clear relation (such as `i--i+4' for $\alpha$-helix) occurs between hydrogen-bonded residues in a $\beta$-sheet. Thus $\alpha$-helices can be considered ``local'' compared to $\beta$-sheets, as the hydrogen-bonding in $\alpha$-helices is formed between nearby residues in the sequence, while in $\beta$-sheets the hydrogen-bonded residues may be far away in the sequence.

\end{panel}

\subsection{$\beta$-strands}
A $\beta$-strand is a stretch of amino acids with the backbone in an extended configuration, typically 3 to 10 amino acids long. Two or more $\beta$-strands together make up a $\beta$-sheet. Hydrogen bonding patterns in the $\beta$-sheet are distinctly different from those in the $\alpha$-helix. In the $\beta$-sheet, hydrogen bonds occur `laterally' between adjacent strands.

When two $\beta$-strands have the same direction, they are referred to as parallel $\beta$-sheets. Conversely, if the two $\beta$-strands have opposite directions, they are known as antiparallel $\beta$-sheets. These two types of $\beta$-sheets can be distinguished by the geometry of the hydrogen bonds. In parallel $\beta$-sheets, the hydrogen bonds are formed diagonally between the carbonyl group of one amino acid residue and the amide group of the neighboring residue on the adjacent strand, resulting in a slanted hydrogen bonding pattern. On the other hand, in antiparallel $\beta$-sheets, the hydrogen bonds are formed directly between the carbonyl group of one amino acid residue and the amide group of the neighboring residue on the opposite strand, creating a linear hydrogen bonding pattern \figref{par_vs_antipar}.
Please refer to the \panelref{ChIntroPS:sheets} below for further details on beta-sheets.

\subsection{Loops}

In loop regions, see also \figref[c]{ChIntroPS-ss-detail}, there is no regular pattern of hydrogen bonding. Nevertheless, the hydrogen bond donors and acceptors, as present in the backbone, do need to form hydrogen bonds. In loop structures, the backbone atoms may make hydrogen bonds with the solvent, with the sidechains of polar amino acids, or even with backbone atoms -- but not in a regular pattern.

The configuration of loops are much less regular, or ordered compared the helices and $\beta$-sheets. Generally, loops lie on the surface of a protein, and are much more solvent exposed. Often loop regions can be flexible, and can change conformation in the functional state of the protein, even when the protein is fully folded.
Loop regions are therefore also much more likely to loose (deletion) or gain (insertion) small parts during evolution. In a (multiple) sequence alignment, loop regions typically contain many gaps compared to helical or sheet regions.
Very long loops ($> 20$ residues) are also called \emph{disordered regions}. Such regions will not take up a rigid three-dimensional structure in their folded state, see also \panelref{ChIntroPS:atypical-ss}.

\begin{figure}
\centerline{\figlab{A}\raisebox{-0.9\height}{
\includegraphics[height=0.3\linewidth]{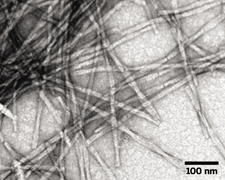}}
\figlab{B}\raisebox{-0.9\height}{
\includegraphics[height=0.3\linewidth]{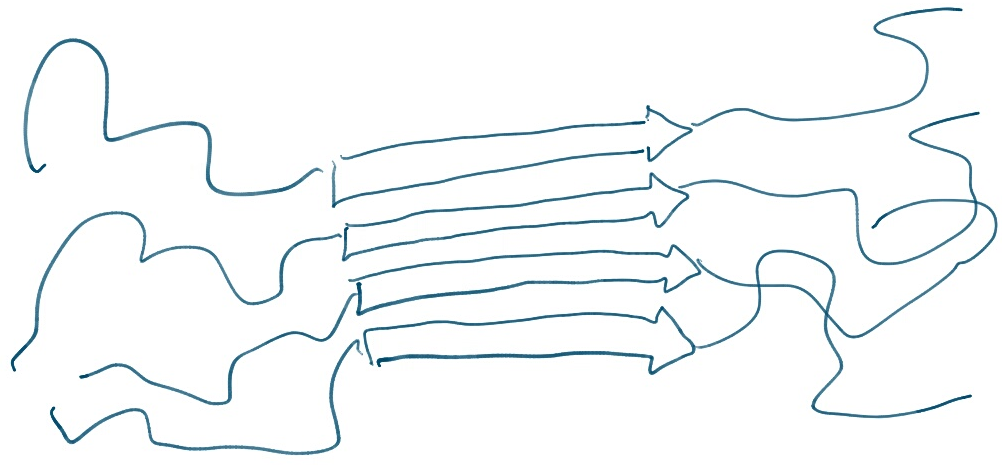}}}
\vspace*{2ex}
\centerline{\figlab{C}\raisebox{-0.9\height}{
\includegraphics[height=0.35\linewidth]{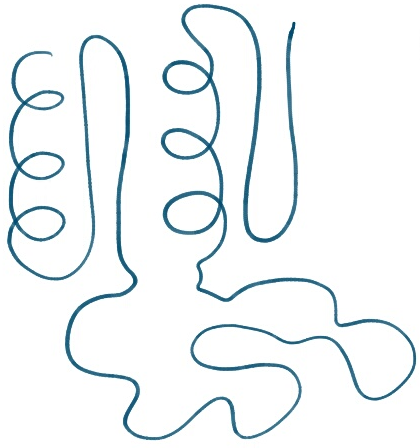}}
\figlab{D}\raisebox{-0.9\height}{
\includegraphics[height=0.35\linewidth]{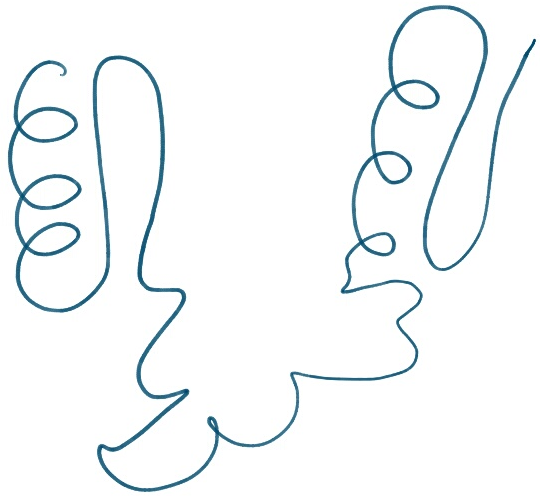}}}
\vspace*{2ex}
\centerline{\figlab{E}\raisebox{-0.9\height}{
\includegraphics[width=0.95\linewidth]{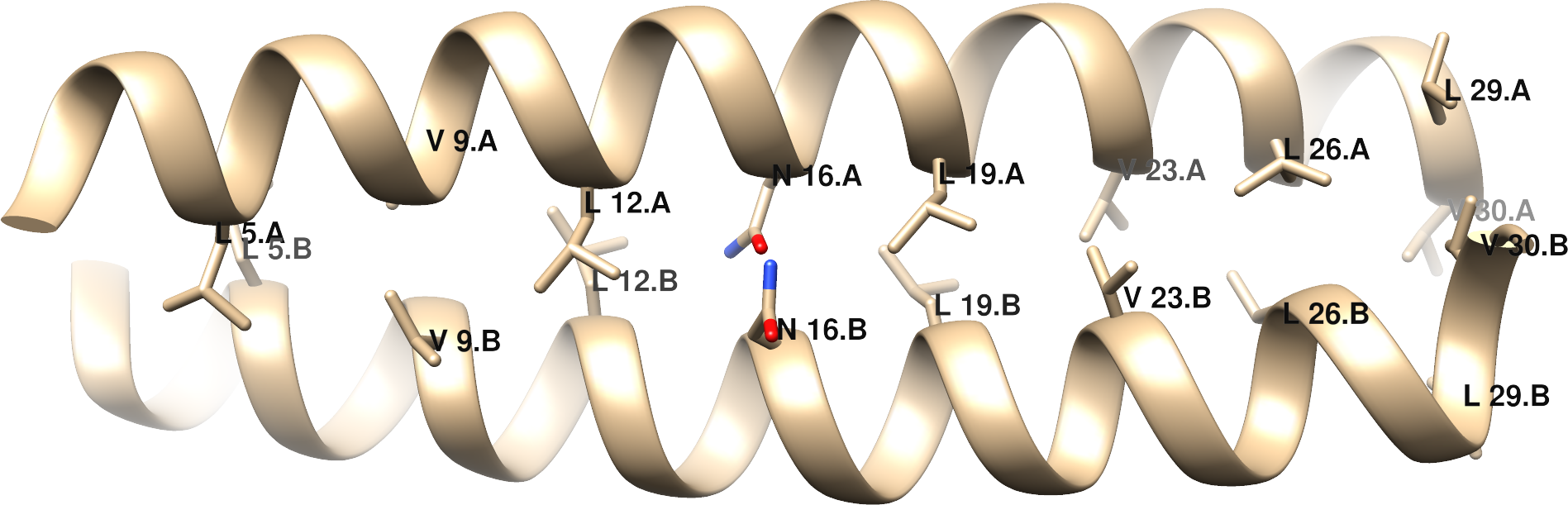}}}
\vspace*{3ex}
\centerline{\figlab{F}\raisebox{-0.9\height}{
\includegraphics[width=0.45\linewidth]{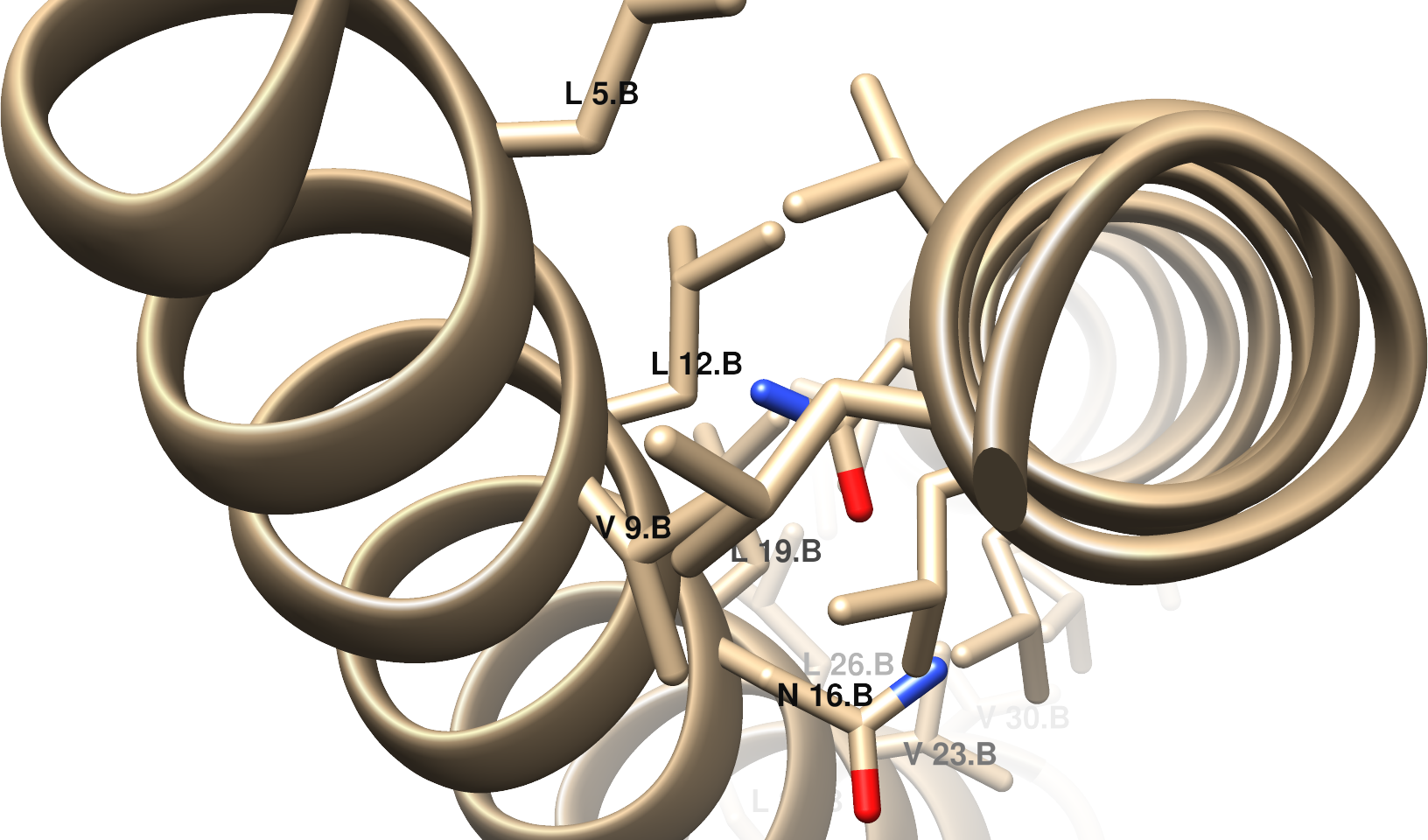}}
\figlab{G}\raisebox{-0.9\height}{
\includegraphics[width=0.5\linewidth]{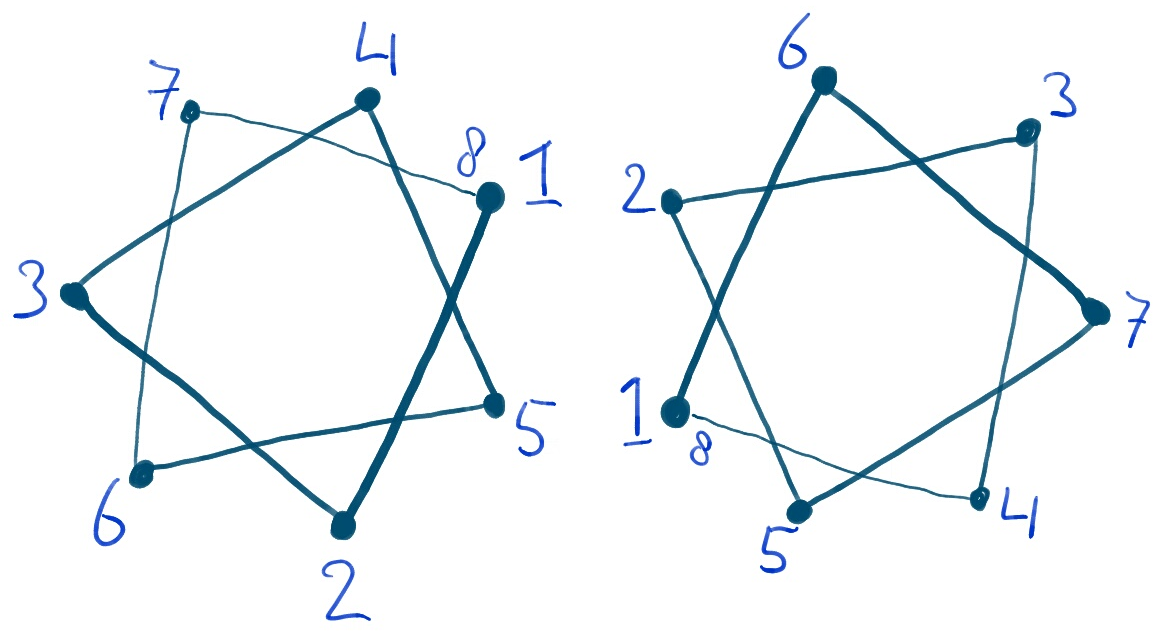}}}
\caption{Three main classes of a-typical protein structures: 
amyloid-fibrils or $\beta$-fibrils (A), as an example of bound ordered structure flanked by disordered loops or termini (B). 
Disordered proteins or regions; shown are schematically a disordered loop within a protein domain (C) and a disordered linker between two protein domains (D). 
(E+F) $\alpha$-helical coiled coils (\pdbref{2ZTA}) which are characterized by the repetition of a Leucine every 7th residue, hence also referred to as `leucine zippers'. Shown length-wise (E), from the top (F) and schematically (G). } \label{fig:ChIntroPS:atypical-ss}
\end{figure}

\begin{bgreading}[Unusual secondary structures]
\label{panel:ChIntroPS:atypical-ss}

In addition to `typical' proteins secondary structures, there are three main classes of `a-typical' cases: amyloid fibrils, coiled-coils, and disordered proteins. Examples are shown in \figref{ChIntroPS:atypical-ss}.
\begin{compactitem}
\item[\bf Amyloid fibrils] ($\beta$-fibrils) are a particular case of $\beta$-sheets. Here the $\beta$-strands are also formed between the chains of different protein molecules, and such structures can become infinitely long. The resulting fibrils may form larger aggregate fibers, that may disrupt the cell functioning or even kill cells. Initially, the ability to form these fibrils was thought to be a particular property of specific proteins and associated with particular pathologies, like the prion protein in scrapie (sheep), mad cow disease (cattle) or Creutzfeldt-Jacob's disease (human). It has now become clear that the ability of proteins to form amyloidic structures tends to be generic \cite{Dobson2003}. Several other diseases have now been associated with the formation of `amyloid plaques', which are large-scale deposits of $\beta$-fibrils that can be highly disruptive to tissue. A well-known example is some forms of Parkinson's disease. However, it is not yet clear in most cases if these plaques are involved in causing the disease or merely a result of the disease process.

\item[\bf The coiled-coil] is a twisted rod formed by a pair of $\alpha$-helices, this is shown in \figref[b]{ChIntroPS:atypical-ss}. It resembles a pair of tweezers, with one end slightly open, and both helix ends binding on either side of the DNA double helix in certain DNA binding proteins. The coiled-coil has a repetitive element of 7 residues where both helices are in direct contact. Typically, every 7th residue is a leucine, and valines or isoleucines are found in between. This creates a pattern like \texttt{Lxx[VI][VI]xxL} (\texttt{[VI]} means either V or I at that position). These structures are also referred to as ``leucine zippers''and ``leucine-rich repeats''because of the repeating leucine every 7th residue.

\item[\bf Disordered protein regions:] some proteins never fold in a fixed three-dimensional structure, and are referred to as  \textbf{``disordered proteins''}. These lack a folded structure, but display a highly flexible, random-coil-like conformation under physiological conditions. They will be further discussed in \chref[nn]{ChSSPred} will briefly go into prediction of disordered proteins and regions based on sequence patterns. Many proteins contain large disordered segments (33\% of eukaryotic, 2\% for archaea, and 4.2\% in bacteria) \cite{Ward2004}.

\end{compactitem}

\end{bgreading}

\subsection{Phi and psi angles}
\label{sec:ChIntroPS:phi-psi}

\begin{figure}
\centerline{
\includegraphics[height=0.75\linewidth]{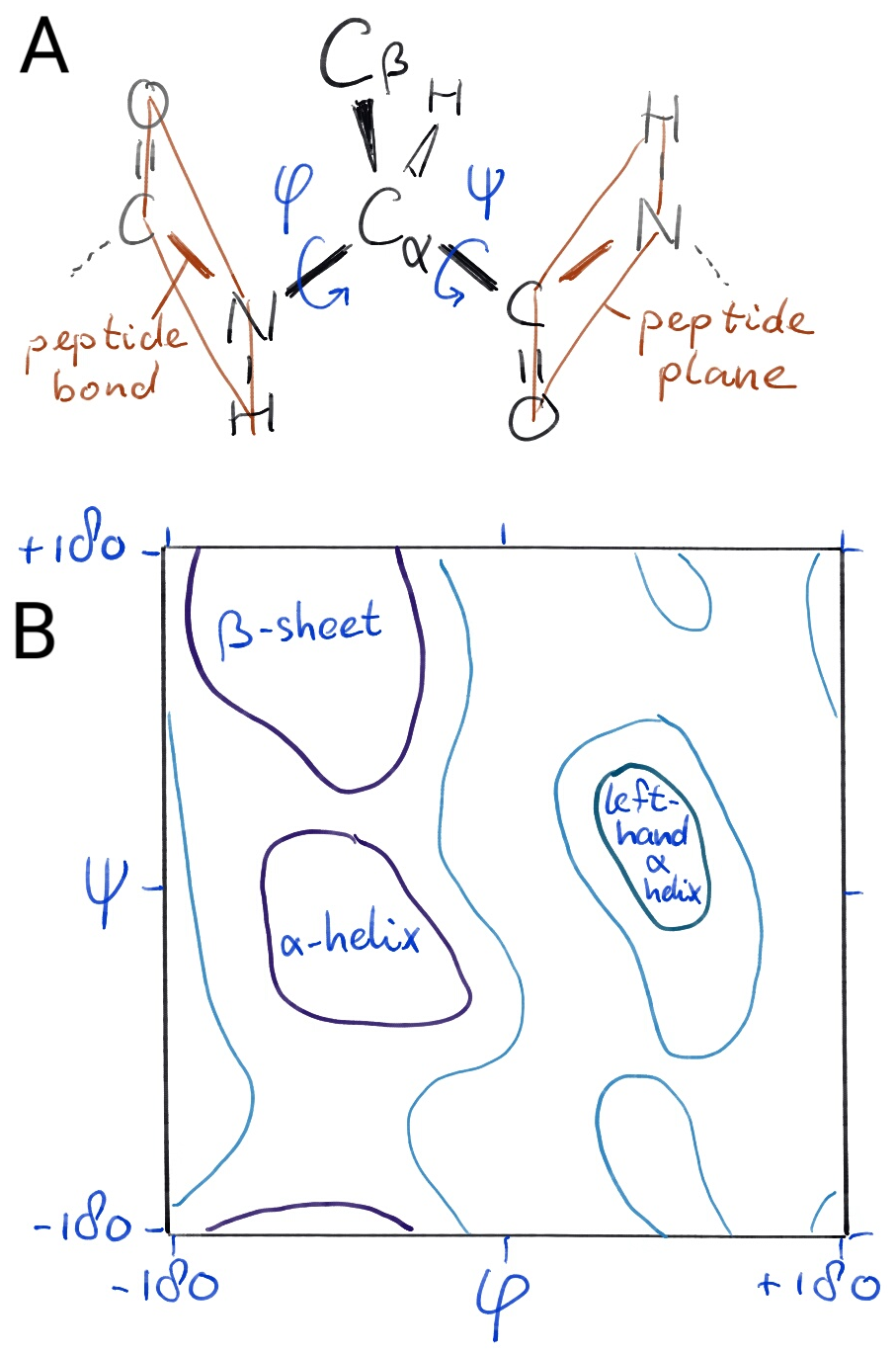}
\includegraphics[height=0.75\linewidth]{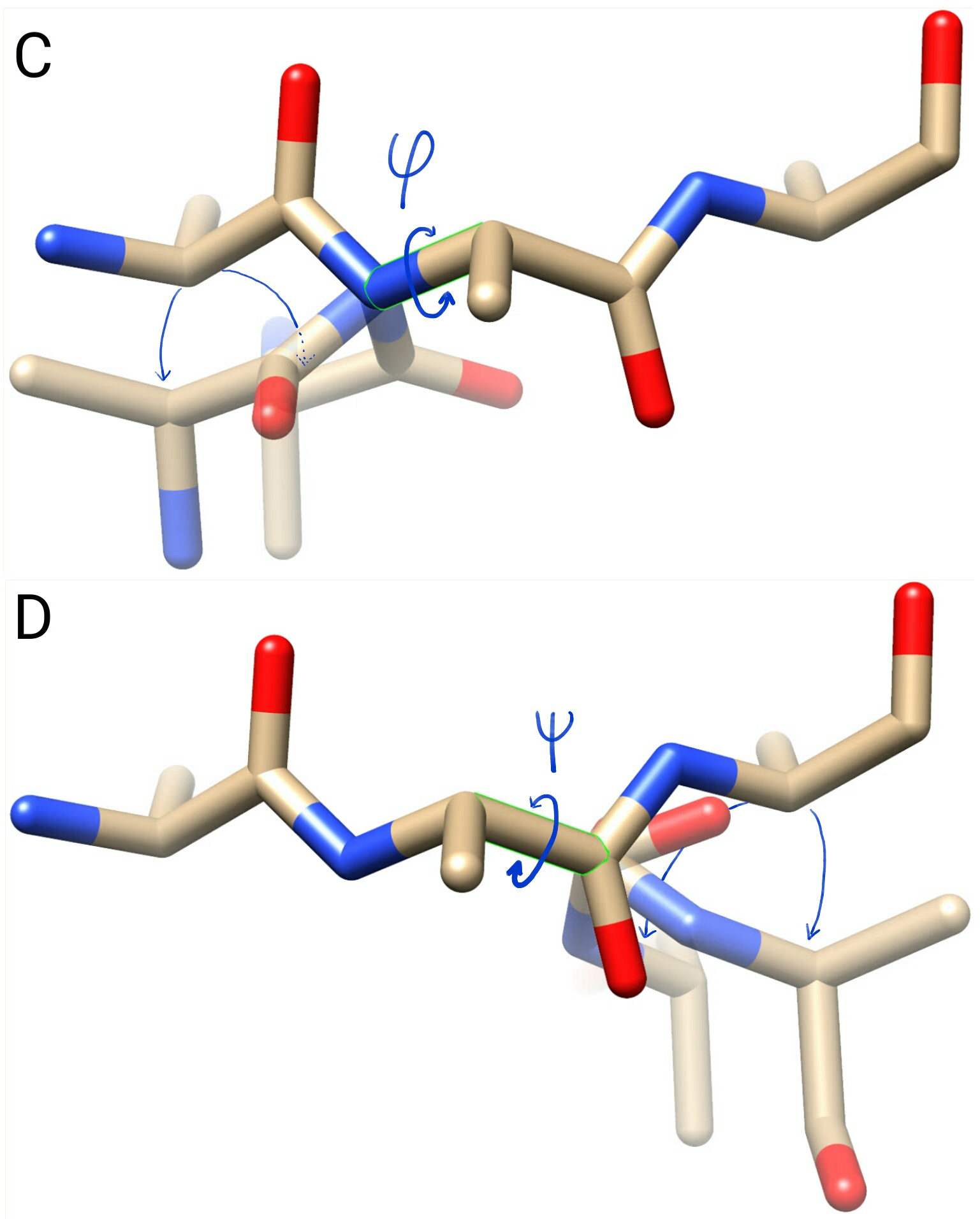}}
\caption{(a) Phi or $\Phi$ and psi or $\Psi$ angles defined in the backbone of an Alanine. (b) Phi and psi angles analyzed in a ramachandran plot. The contoured areas indicated allowed (light) and preferred (dark) combinations of phi and psi angles, which coincides with the two major secondary structure elements: $\beta$-sheets and $\alpha$-helices; in addition the smaller area of left-handed $\alpha$-helices can be seen in the positive quadrant~\cite{Chen2010}. (c+d) Backbone re-arrangements in a tri-Alanine peptide, when adjusting the $\Phi$ backbone dihedral angle of the central residue (c) or the $\Psi$ angle (d). The backbone nitrogen (N) is in blue, the carboxyl (C=O) oxygen in red, and the carbon atoms in tan. The C$\alpha$ atom is the one without a (red) oxygen bound, and with the C$\beta$ (tan) branching off of the backbone. These structural illustrations were created using  Chimera.} \label{fig:ChIntroPS-phi-psi} 
\label{fig:ChIntroPS-rama}
\end{figure}

The backbone of a peptide consists of two flexible chemical bonds: NH--C$\alpha$ and C$\alpha$--CO. These bonds can rotate around their axes, and they are referred to as torsion angles or dihedral angles. In this context, we will use the term torsion angles. The NH--C$\alpha$ torsion angle is denoted as $\Phi$ or phi and is located at the beginning of each residue. The C$\alpha$-CO torsion angle is denoted as $\Psi$ or psi and is found at the end of each residue. \figref{ChIntroPS-phi-psi} illustrates this arrangement. The omega torsion angle is covered in the \panelref{ChIntroPS:omega}.

Specific combinations of phi and psi angles allow the formation of favourable amide (NH) to carbonyl (C=O) hydrogen bonding patterns in the backbone. These combinations facilitate the formation of either $\alpha$-helix or $\beta$-sheet structures. However, certain phi and psi angle combinations can result in clashes within the backbone or between adjacent sidechains in the protein chain, particularly if the sidechains are large. In some cases, clashes can even occur between the larger sidechains and the backbone of neighboring residues. Steric hindrance caused by both the backbone atoms and sidechains restrict the occurrence of certain combinations of phi and psi angles. Consequently, the number of potential conformations that may be adopted by the polypeptide is reduced. Sequence and propensity patterns that arise from this are exploited in secondary structure prediction, to which we will turn in \chref{ChSSPred}.

Based on known protein structures, we can derive empirical distributions of phi and psi angle combinations. This distribution is visualised in a so-called Ramachandran plot of phi (horizontal) vs. psi (vertical), as shown in \figref{ChIntroPS-rama}. Firstly, we can observe that only some combinations of phi and psi angles are allowed, e.g.\@ light and dark outlined areas, whilst others are very uncommon (outside areas). The allowed region are the secondary structure elements, such as the $\alpha$-helix (with negative phi and psi angles) and $\beta$-sheet (with negative phi and positive psi angles). Additionally, there is a smaller area that corresponds to the left-handed helix, which is observed but less frequently encountered. The areas without data points, the disallowed regions, indicate combinations of phi and psi angles that result in steric hindrance among the backbone atoms and are therefor not observed.

\begin{bgreading}[The omega torsion angle]
\label{panel:ChIntroPS:omega}
The peptide bond (between C=O and N--H) chemically connects two amino acid residues together. The consecutive peptide bonds form the backbone of the protein. Strictly speaking the peptide bond is a torsion angle like the phi and psi angles, however this bond is different. Due to the physiochemical properties of this bond it cannot rotate freely, this is related to the fact that it is in between a C=O and N--H group. This bond angle is called $\omega$ or omega. In proline residues, it can switch between two possible angles in a process called `proline isomerisation'.
\end{bgreading}

\subsection{Secondary structure assignment}
Secondary structure \emph{assignment} involves determining the secondary structure class for each residue in a protein based on its structure. It is a structure-based definition for secondary structure. Protein structures are typically stored as a set of coordinates for each atom in the structure, see also \chref{ChDBClass}. 
Various features such as phi and psi angles or hydrogen bonding patterns, can be used to assign secondary structure.

The most commonly used method to assign secondary structure is the Dictionary of Secondary Structure of Proteins, or DSSP \cite{Kabsch1983}, but several others exist such as Stride \cite{Heinig2004}. DSSP first assigns hydrogen bonds to pairs of atoms, and uses these pairs to infer the secondary structure. For example, if several consecutive residues have hydrogen bonds that are four places ahead in the sequence, these residues are designated to be part of an $\alpha$-helix by DSSP. Minimum lenghts of secondary structure elements are also considered by these methods, to avoid assigning a single residue as an $\alpha$-helix.

\section{Tertiary structure}

\begin{figure}
\centerline{
\figlab{A}
\includegraphics[width=0.45\linewidth]{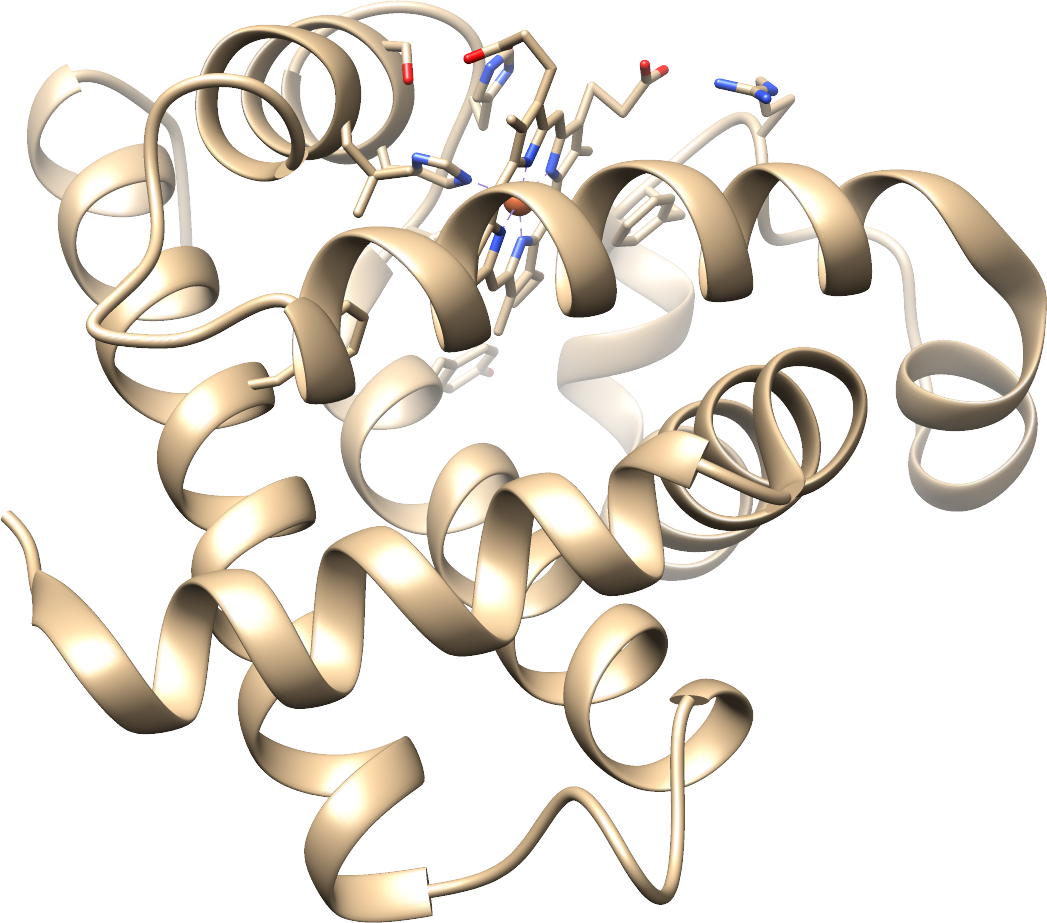}
\figlab{B}
\includegraphics[width=0.45\linewidth]{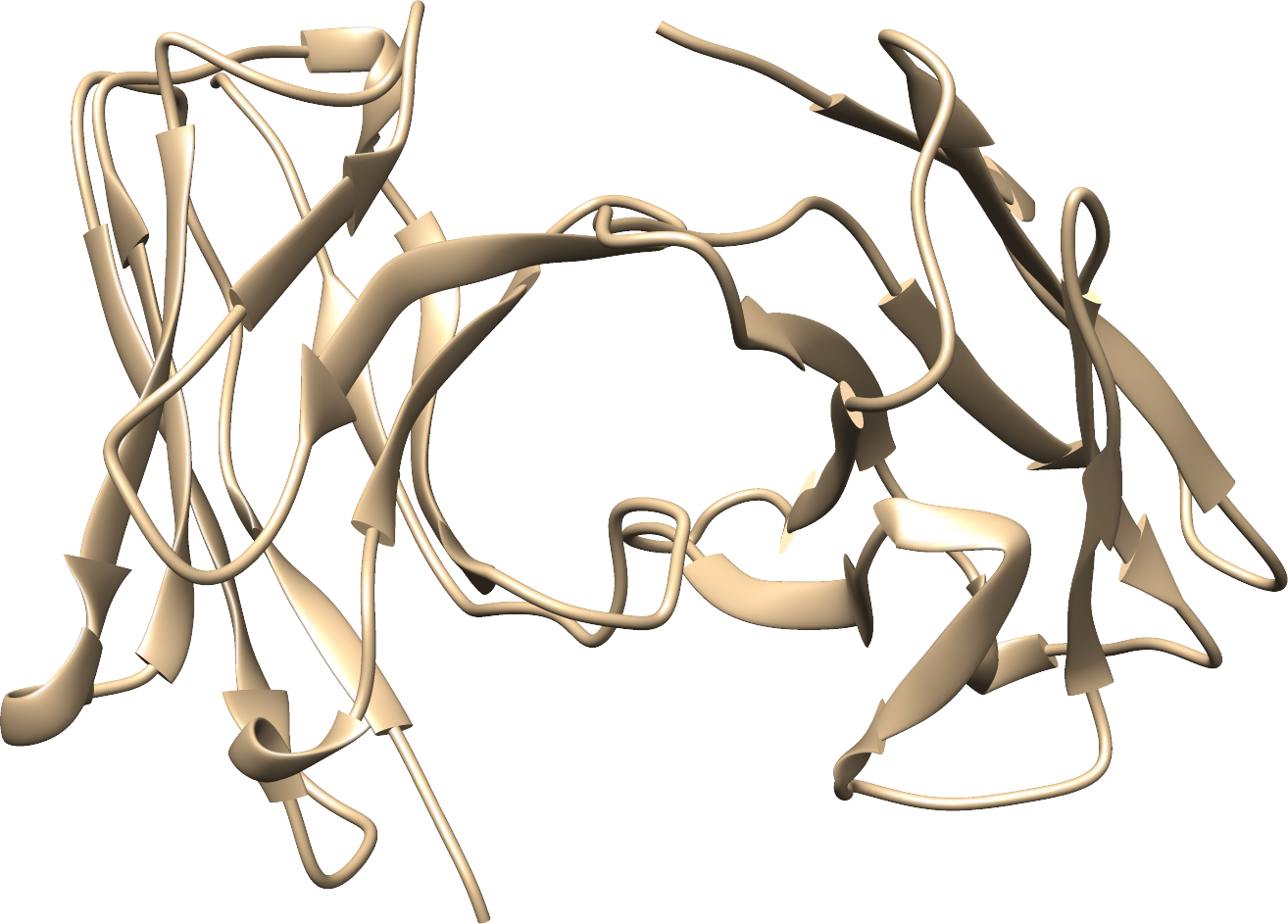}
}\vspace*{2ex}\centerline{
\figlab{C}
\includegraphics[width=0.45\linewidth]{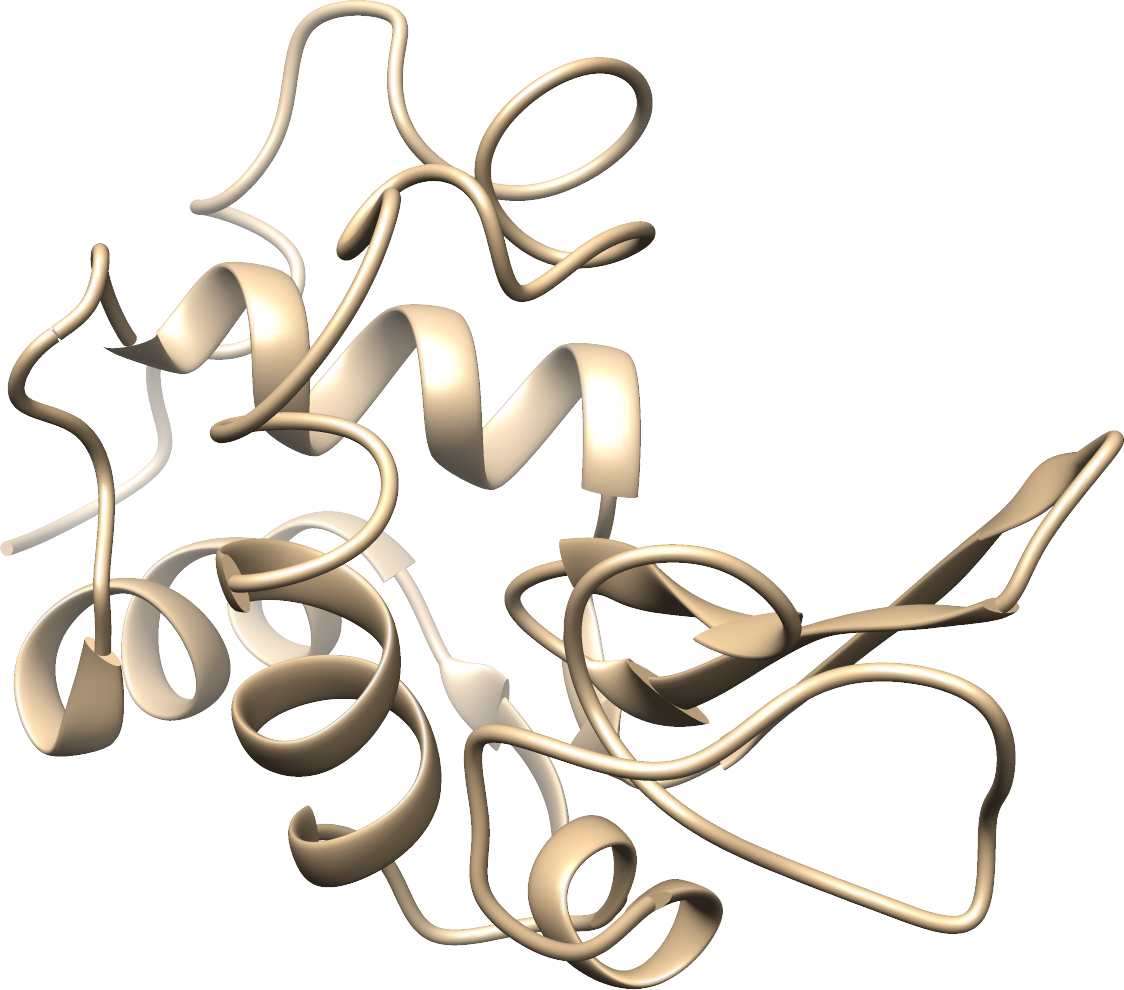}
\figlab{D}
\includegraphics[width=0.45\linewidth]{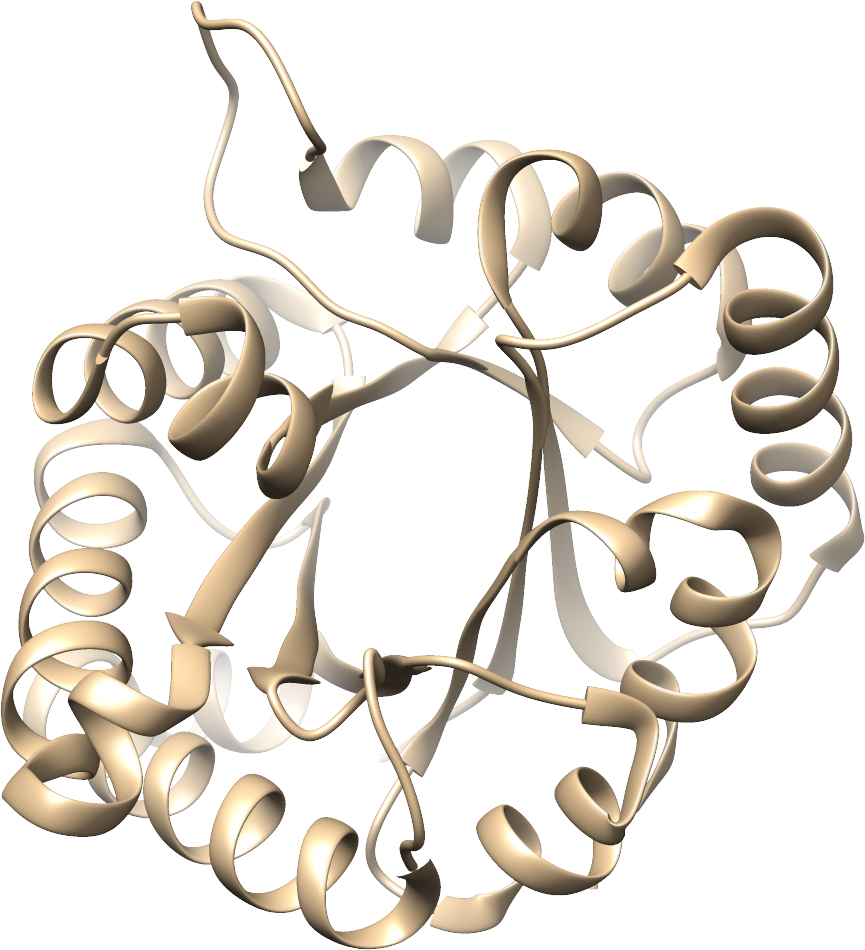}
}
\caption{The four main protein fold classes, here showing a more or less famous example for each of them. 
(A) all-$\alpha$: myoglobin, solved in 1960 by Sir John Kendrew, for which he received the Nobel prize \cite[\pdbref{1mbn},][]{Kendrew1960}, consisting of only alpha-helices. 
(B) all-$\beta$: Immunoglobulin domain, consisting of only beta-strands \cite[\pdbref{1igt},][]{Harris1997}.
(C) $\alpha$/$\beta$ lysozyme by DC Phillips, where one domain is helical, and another strands (here on the bottom right) \cite[\pdbref{1lyz},][]{Diamond1974}.
(D) $\alpha$+$\beta$ triose phosphate isomerase, where helical and strand regions intermingle \cite[\pdbref{1tim},][]{Banner1976}.
} \label{fig:ChIntroPS-fold-classes}
\end{figure}

The the tertiary structure of the protein, which represents its complete structure, consists of secondary structure elements, or motifs (see also \panelref{ChIntroPS:ss-motifs}). The arrangement of secondary structure elements along the protein chain, and their folding to establish contacts in the three-dimensional structure, is called the \emph{protein fold}. This is also referred to as `protein topology'. Four main fold categories are distinguished: all-$\alpha$, all-$\beta$, $\alpha$/$\beta$ and $\alpha+\beta$ as shown in \figref{ChIntroPS-fold-classes}.

\begin{bgreading}[Secondary structure motifs]
\label{panel:ChIntroPS:ss-motifs}
Secondary structure elements often occur in the protein structure in particular combinations, called `secondary structure motifs'. Some of these can be very informative, e.g.\@ for assessing protein function, which is why they form the basis of structural classification schemes. As an example, \figref{ChIntroPS-fold-classes} shows the four main fold classes from one of the protein databases, SCOP \cite{Andreeva2008}. For a comprehensive treatise on secondary structure motifs, we refer to \citet{BrandenTooze}. Also, you can find more on this in \chref[nn]{ChDBClass}.

\end{bgreading}

\subsection{Hydrophobic core}
\label{sec:ChIntroPS:hydrophobic-core}
Proteins in a cell are typically surrounded by water (except for transmembrane proteins which are located inside of cell membranes). Water is a polar molecule: with the oxygen atom slightly negative and the hydrogen slightly positive, see also  \figref{ChIntroPS-hbonds}. 
Hydrophobic sidechains cannot make hydrogen bonds with the water molecules, therefore the solvent avoids to make contact with hydrophobic residues. Oil is also a hydrophobic substance, and as you may well know: oil and water do not mix well.

In a protein, the hydrophobic residues want to be shielded from the water, in any stable configuration. The result is that the protein will adopt a conformation in which the exposure of the hydrophobic sidechains to the water is minimized; hydrophobic residues will tend to become buried in the interior of the protein. This effect is known as the `hydrophobic effect' and is the main driving force for protein folding. In the folding process, this leads to what is called the `hydrophobic collapse'. \figref{ChIntro-collapse} sketches the role of hydrophobic residues in the folded and unfolded states of a protein. We will come back to the folding process and the thermodynamics and driving forces behind in much more depth in \chref[nn]{ChIntroDyn}.

For now, it is important to realize that the interior of the protein will consist mostly of residues with hydrophobic sidechains, but the backbone is polar. It is not possible to keep all the polar backbone parts of the buried hydrophobic residues at the surface of the protein. At the surface, the polar sidechains as well as the backbone, form hydrogen bonds with the water, but the backbone of the buried hydrophobic residues cannot do this. This creates a problem, as it is very unfavourable if these backbone hydrogen bonding capacity remains unsatisfied. The solution to this problem is the formation of secondary structure, as covered in the previous section. Regular secondary structures will therefore usually make up the core of the protein.

\subsection{Protein domains}
Protein domains are conserved regions that will be mentioned often in this book. 
More generally, we could define them as self-folding, evolutionary conserved subunits of structure. Domains typically have a specific molecular function, and may recurrently appear in different proteins. Most eukaryotic proteins have multiple domains, which may be linked together by a small linker, or large disordered regions. There are serveral distinct ways in which domains may be described, each of which will be explained in further detail elsewhere in this book. In \chref{ChDBClass} we will see how structural domains can be defined; in \chref{ChIntroPred} it becomes clear that structure prediction is most effective at the domain level.

\section{Quarternairy structure}
Protein-protein interactions (PPIs) involve the binding of two or more proteins, leading to the formation of a protein complex known as the quaternary protein structure. This structure represents a natural extension of the primary, secondary, and tertiary structures. It is worth noting that protein function often emerges at the level of the quaternary structure, as it determines the specific function performed by the protein complex. We will come back to protein function, and the role of interactions, in \chref{ChFuncPred}.

\section{Key points}
\begin{compactitem}
\item Proteins fold from an unstructured polypeptide coming from the ribosome into their functional native conformations.
\item Structure Basics: 
\begin{compactitem}
\item primary, secondary, tertiary, tertiary
\item phi/psi angles
\item hydrogen-bonds
\end{compactitem}
\item Loops tend to be more flexible
\item Hydrogen bonds may be satisfied by backbone, sidechain or water
\item PDB \& Structural genomics: bias in data
\item Protein structure may be predicted from sequence
\item Function may be derived from structure
\end{compactitem}

\section{Further Reading}
\begin{compactitem}
\item ``Sequence Analysis'' \cite{Durbin}
\item ``Introduction to Protein Structure'' \cite{BrandenTooze}
\end{compactitem}

\section*{Author contributions}
{\renewcommand{\arraystretch}{1}
\begin{tabular}{@{}ll}
\ACtxt: &   AJ, EvD, BS, HM, KAF, SA \\
\ACfig: &   HM, JG, KAF, SA \\
\ACref: &   HM, KAF, SA \\
\ACproof:&  BS, KAF, SA \\
\ACeds: &   SA, KAF
\end{tabular}}

\noindent
The authors thank \RH~\RHid{} and \NB~\NBid{} for non-expert feedback.

\mychapbib

\clearpage
%\addcontentsline{toc}{chapter}{\bibname}
%\bibliography{strucbioinf}

\cleardoublepage

\end{document}